\documentclass[a4paper,twocolumn,floatfix,longbibliography]{revtex4-1}

\usepackage{graphicx}
\usepackage{siunitx}
\usepackage{pifont}
\usepackage{MnSymbol}
\usepackage{amsmath}
\usepackage{xcolor}
\usepackage{braket}
\usepackage{hyperref}
\usepackage[english]{babel}

\begin{document}

\title{\texorpdfstring{High-resolution nanosecond spectroscopy of even-parity Rydberg excitons in Cu$_{2}$O}{High-resolution nanosecond spectroscopy of even-parity Rydberg excitons in Cu2O}} 

\author{Joshua P. Rogers, Liam A. P. Gallagher, Danielle Pizzey, Jon D. Pritchett, Charles S. Adams, Matthew P. A. Jones}

\affiliation{Department of Physics, Durham University, Durham DH1 3LE, United Kingdom}

\author{Chris Hodges, Wolfgang Langbein, Stephen A. Lynch}
\affiliation{School of Physics and Astronomy, Cardiff University, Cardiff CF24 3AA, United Kingdom}

\date{\today}

\begin{abstract}

We present a study of even-parity Rydberg exciton states in cuprous oxide using time-resolved second harmonic generation (SHG). Excitonic states with principal quantum number $n=5-12$ were excited by nanosecond pulses around 1143~\si{\nano\meter}. Using time-resolved single-photon counting, the coherently generated second harmonic was isolated both temporally and spectroscopically from inelastic emission due to lower-lying free and bound excitonic states, which included narrow resonances at 1.993~\si{\electronvolt} associated with an long lifetime of $641\pm7 ~\si{\micro\second}$. The near transform-limited excitation bandwidth enabled high-resolution measurements of the exciton lineshape and position, from which we obtained values for the quantum defects of the S and D excitonic states associated with the appropriate crystal symmetries. Odd-parity P and F excitonic states were also observed, in accordance with predicted quadrupole-allowed two-photon excitation processes.   We compared our measurements to conventional one-photon spectroscopy in the same sample, and find that the SHG spectrum is cut off at a lower principal quantum number ($n=12$ vs $n=15$). We attribute this effect to a combination of spatial inhomogeneities and local heating, and discuss the prospects for observing higher principal quantum number even-parity states in future experiments.
\end{abstract}


\maketitle

\section{Introduction}

The observation of high principal quantum number Rydberg excitons in cuprous oxide~\cite{Kazimierczuk2014} (Cu$_{2}$O) has brought about renewed interest~\cite{Thewes2015,Gruenwald2016,Walther2018a,Zhang2018,Heckoetter2020a} in the material. A major motivation is the potential to combine long-range van der Waals interactions (which scale with principal quantum number, $n$, as $n^{11}$) between excitons with the quantum optical properties of exciton-polaritons~\cite{Khazali2017,Walther2018a,Poddubny2019}. This concept of Rydberg quantum optics builds on previous work using Rydberg states of  laser cooled and thermal atomic vapours, where  single-photon sources and photon-photon interactions have been demonstrated~\cite{Firstenberg2016,Adams2019}.

Mott-Wannier excitons were first observed in cuprous oxide by Hayashi \textit{et al.}~\cite{Hayashi1950,Hayashi1952} and Gross \textit{et al.}~\cite{Gross1956}, using absorption spectroscopy~\cite{Hayashi1950,Hayashi1952,Gross1956,Agekyan1977,Kavoulakis1997,Washington1977}. The highest valence and lowest conduction bands arise, respectively, from the 3d and 4s orbitals of the Cu atoms and are the origin of the lowest energy (yellow) exciton series.  Because both bands are of even-parity, excitons with an odd-parity P-type envelope are dipole allowed~\cite{Elliott1957}.  The seminal study by Kazimierczuk \textit{et al.}~\cite{Kazimierczuk2014} resolved excitons up to principal quantum number $n$ = 25 and more recently Versteegh \textit{et al.}~\cite{Versteegh2021} measured up to $n$ = 30.   At these highest observed states, the spatial extent of the excitons is greater than 1~\si{\micro\meter} and the associated dipole moment is large enough to exhibit evidence of a Rydberg blockade~\cite{Kazimierczuk2014,Walther2018,Heckoetter2018,Heckoetter2020} arising from the van der Waals interactions between the excitons.

\begin{figure}
    \centering
    \includegraphics[width=68mm]{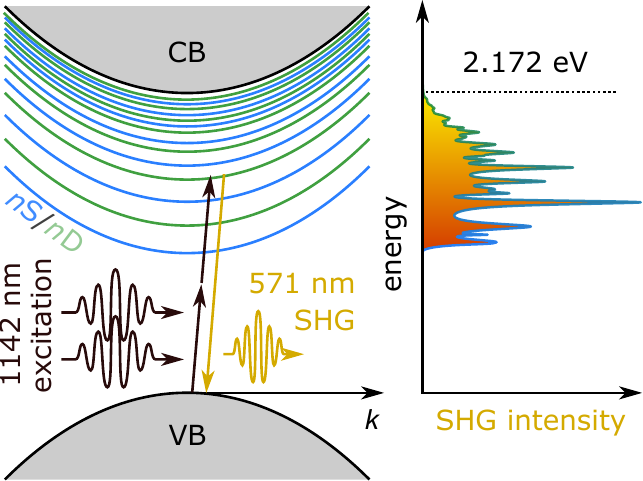}
    \caption{The yellow exciton series of Cu$_{2}$O arises from the highest valence band (VB) and lowest conduction band (CB).  Under two-photon excitation (TPE) and along certain crystallographic axes, Cu$_{2}$O permits second-harmonic generation (SHG).  SHG is enhanced when the combined energy of the two excitation photons is resonant with an even-parity exciton, predominantly with S (blue parabolas) or D (green parabolas) envelope.  SHG is suppressed above the bandgap.  The SHG spectrum on the right, which is shown in more detail in Fig.\,\ref{Fig:SpectralFigure}, is acquired by scanning the excitation energy over the series.}
    \label{Fig:ExcitonCartoon}
\end{figure}

The first study of even-parity excitons in cuprous oxide up to $n=5$ was made by Fr\"{o}hlich \textit{et al.}~\cite{Froehlich1979,Uihlein1981}.  The measured observable in these and other experiments~\cite{Matsumoto1996} was either few percent changes in the intensity of the transmitted laser light or spectrally-resolved photoluminescence (PL).  The resulting two-photon excitation (TPE) spectra consist primarily of even-parity excitons with S- and D-type envelopes as shown in Fig.~\ref{Fig:ExcitonCartoon}.  

From a quantum optics perspective, the study of even-parity excitons offers several advantages over that of the odd-parity states. TPE photoluminescence and SHG provide a signal at approximately twice the frequency of the excitation laser, making it easier to remove stray light. Unlike one-photon absorption, where interactions with optical phonons~\cite{Ueno1969} gives rise to strong background absorption, SHG is a background-free process dominated by the excitonic resonances.  The signal therefore vanishes above the bandgap of the material. These properties were essential to a recent demonstration of the microwave modulation of an optical carrier in cuprous oxide~\cite{Gallagher2021}. It is therefore an important question whether the even-parity spectrum may be resolved up to the same high principal quantum numbers ($n>25$) observed in one-photon excitation of the $n$P series~\cite{Kazimierczuk2014,Versteegh2021}.

Recently, Mund \textit{et al.}~\cite{Mund2018,Mund2019,Farenbruch2020,Rommel2020} have studied even-parity excitonic states up to 9S using second-harmonic generation (SHG). In these experiments~\cite{Mund2018,Mund2019,Farenbruch2020,Rommel2020}, broadband femtosecond pulses simultaneously excited a range of Rydberg states, and the resulting second harmonic was spectrally resolved using a monochromator. Although cuprous oxide possesses inversion symmetry, SHG becomes allowed at exciton resonances when quadrupole terms are considered in the excitation or emission operators. The resulting selection rules are considered in detail in \cite{Schoene2016,Mund2018}. SHG may also occur if the crystal symmetry is lifted either by residual strain~\cite{Mund2019}, lattice imperfections or by externally applied fields~\cite{Rommel2020, Farenbruch2020}.

In this paper we present an novel approach to SHG spectroscopy based on a continuously tuneable diode laser chopped into high-intensity nanosecond pulses, the duration of which can be varied using a fiber-based electro-optic modulator.  The linewidth is near transform-limited to $\sim 50$~\si{\nano\electronvolt} for a typical pulse duration of 50~\si{\nano\second}, which is substantially less than the \si{\micro\electronvolt} spectral width of the excitonic features. This narrow linewidth enables direct measurement of the lineshape of high-$n$ excitons by scanning the laser.  The high excitation intensity required for SHG is provided by a Raman fiber amplifier that achieves an average output power of up to 5~\si{\watt}.  A representation of the excitation scheme is shown in Figure~\ref{Fig:ExcitonCartoon}. We combine the narrowband pulsed excitation with time-resolved single-photon counting detection, which enables measurements of the temporal profile of the emitted light over timescales from nanosecond to milliseconds, with resolution down to 50~\si{\pico\second}.  Here, we show that this reveals the presence of weak decays from long-lived bound excitons states. We also perform spatially-resolved measurements of the SHG intensity, with a resolution on the micron scale.

\section{Experimental overview}

\begin{figure*}
    \centering
    \includegraphics[width=150mm]{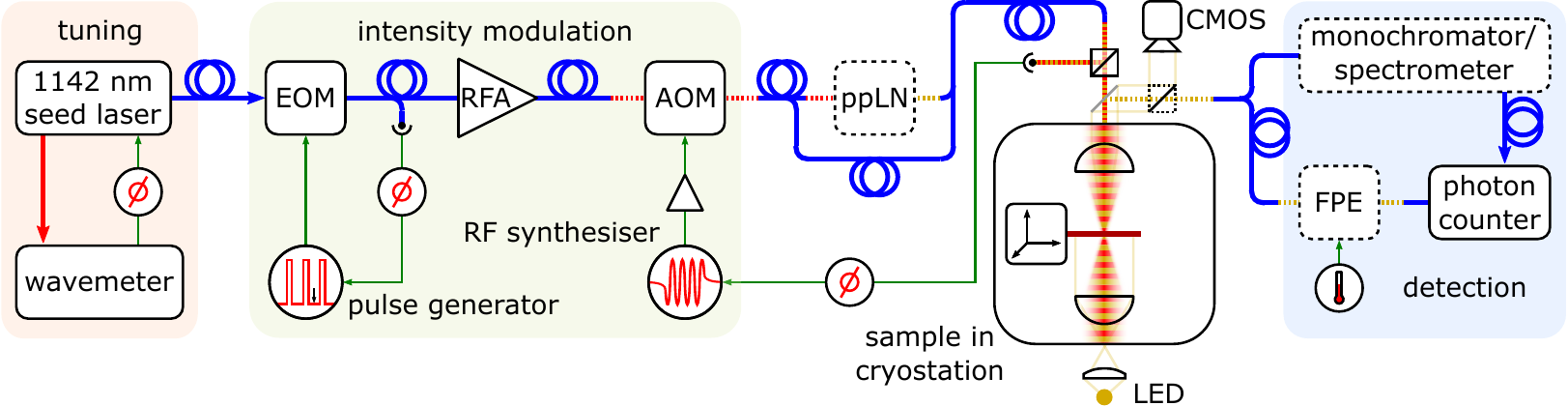}
    \caption{Schematic of the experimental platform. The seed laser is stabilized using a commercial wavemeter and carried primarily by polarization-maintaining single-mode optical fiber (blue lines). After pulse modulation (EOM) and amplification (RFA) the free-space IR beam (dashed red lines) passes through a noise eater/pulse picker (AOM) before transfer to the cryostat, where it passes through in-vacuo aspheres mounted either side of the sample. Backscattered SHG (yellow dashed lines) is separated using a dichroic mirror (gray line) and fiber coupled to the detection system. Also shown are the LED and camera (CMOS) used for imaging the sample. Components in dashed boxes (ppLN crystal, camera flip mirror, and Fabry-P\'{e}rot etalon (FPE)) may be inserted or removed by redirecting appropriate optical fibers. Electrical feedback signals are represented by green arrows.}
    \label{Fig:Setup Figure}
\end{figure*}

The overall experimental schematic is shown in Fig.\,\ref{Fig:Setup Figure}.  All the data presented in this paper were acquired from natural cuprite gemstones from the Tsumeb copper mine in Namibia.  The material was oriented using a Laue camera (Multiwire Labs Ltd MWL120) and cut to expose the (111) plane.  Rectangular samples of $2 \times 3$~\si{\milli\meter} were prepared by slicing the oriented material and polishing it on both sides to a thickness of $100 \pm 10$~\si{\micro\meter}.  Full details of the preparation process are described elsewhere~\cite{Lynch2021}.  

Each lamella of Cu$_{2}$O was positioned in a pocket between two CaF$_{2}$ windows and fixed at one corner with a dab of nail varnish.  The windows were in turn held in a copper mount, also attached by nail varnish for thermal contact.  This mount was fixed to a 3-axis piezo translation stage (Attocube Systems AG nanopositioners: $2 \times $~ANPx101/RES/LT/UHV and $1 \times$~ANPz102/RES/LT/UHV) in a Montana Instruments C2 Cryostation and cooled to a base temperature of 4~\si{\kelvin}.  Two in vacuo aspheric lenses (NA = 0.6; Lightpath Technologies 354105) mounted on the heat-shield and cooled to about 30~\si{\kelvin} were aligned coaxially either side of the sample.  The front lens focused laser excitation onto the sample and collected the back-scattered signal.  The rear lens re-collimated the intense excitation beam to be dumped safely.  Just before the cryostat, a long-pass dichroic filter with cutoff at 785~\si{\nano\meter} (Semrock DI03-R785-T3-25-D) separated signal light from the infrared excitation beam path. The excitation spot size inside the sample was $\sim 0.5~\si{\micro\meter}$. Due to uncompensated birefringence in the CaF$_{2}$ windows the light polarization was slightly elliptical, and its orientation with respect to the in-plane crystallographic axes was not set.

The seed laser (orange box in Fig.\,\ref{Fig:Setup Figure}) was a home-built Littrow-configuration external cavity diode laser (ECDL) based on a ``butterfly'' packaged gain module (Innolume GmbH GM-1140-120-PM-130) centred at 1140~\si{\nano\meter}. The waveguide-type gain module had two output ports. The main output was via a pre-aligned, single mode polarization maintaining fiber pigtail. Placing a 1200~\si{\per\milli\meter} grating at the second free-space output port converted the gain module into an ECDL. By optimizing the geometry of the lever arm holding the grating we designed the ECDL cavity to reach the synchronous tuning regime~\cite{McNicholl1985,Nilse1999,Labachelerie1993}, where the effects of changing the grating angle and the cavity length are appropriately matched to maximize the mode-hop-free tuning range.  This architecture, in conjunction with the anti-reflection coated free-space facet enabled mode-hop-free tuning over a \si{\milli\electronvolt} range. An additional IR etalon (not shown in Fig.~\ref{Fig:Setup Figure}) was used to estimate the short-term linewidth via the side-of-fringe technique, yielding an RMS width of 8~\si{\nano\electronvolt} measured over 500 \si{\milli\second}. 

The zeroth order light at the free space output was passed through two consecutive $> 30$~\si{\decibel} free-space isolators (Thorlabs Inc. IO-4-1150-VLP) and sent to a precision wavemeter (Bristol Instruments 671A-NIR). The wavelength was measured 4 times per second, and computer-controlled tuning and wavelength stabilization was executed by two piezo devices regulating the laser cavity. Firstly, a slip-stick piezo actuator provided coarse tuning,  by moving the grating lever arm in discrete steps corresponding to $\sim 400$~\si{\nano\electronvolt} in excitation energy.  The total coarse tuning range exceeded 1130~--~1170~\si{\nano\meter}.  Scans were performed by incrementing the coarse stepper and measuring the wavelength continuously as the signal was acquired.  Secondly, a piezo stack at the tip of the coarse tuning actuator provided $\pm 40$~\si{\micro\electronvolt} of continuous fine tuning for a given actuator position.  This piezo stack was powered by a 0 - 100~\si{\volt} driver with a resolution of 10~\si{\milli\volt}, which translates to spectral increments of $\sim 4$~\si{\nano\electronvolt}.  The fine tuning was used primarily in conjunction with a feedback loop to stabilize the laser energy to within $\pm 80$~\si{\nano\electronvolt} of the set point.

The main fiber-pigtailed output ($\sim 80$~\si{\milli\watt}) was passed through an in-line fiber optical isolator (Innolume GmbH PMOI-1150-SS) and then chopped into nearly transform-limited pulses by an electro-optic intensity modulator (EOM; EOSpace lithium niobate Mach-Zehnder interferometer) driven by a pulse generator (Hewlett Packard 8131A).  This is shown in the green panel in Fig.\,\ref{Fig:Setup Figure}.  The pulse generator operated at variable repetition rate up to 100~\si{\mega\hertz} and pulse duration down to 700~\si{\pico\second}.  In order to achieve the intensities required for TPE, the pulses were amplified in a Raman fiber amplifier (RFA; MPB Communications Inc RFL-P-5-1143-SF-NS) up to a maximum of 5~\si{\watt} average power. Provided the seed laser delivered $> 0.3$~\si{\milli\watt} of power and the repetition rate was $> 1$~\si{\mega\hertz}, the average output power of the RFA was relatively insensitive to the duty cycle imposed by the EOM.  A pulse duty cycle of 1\% therefore yielded peak amplified powers of several hundred watts~\cite{Dingjan2006}. 

The output of the amplifier passed through a free-space acousto-optic modulator (AOM; Isomet M1080-T80L-1.5) that served two purposes.  For most experiments it was used to stabilize the intensity of excitation light on the sample, as part of a servo loop based on the signal from a photodiode monitoring a pickoff near the sample.  However, for experiments requiring a low duty cycle, for example to measure the lifetime of long-lived states, the minimum average seed power required by the RFA was a limitation. To overcome this, the AOM was also used as a pulse picker, selecting individual pulses or groups of pulses from the pulse train generated by the EOM.  

Overall, the laser system was able to provide an optical power up to 5~\si{\watt} average, with arbitrary pulse duration (CW to 700~\si{\pico\second}) and repetition rate (single shot to 100~\si{\mega\hertz}). The laser could be tuned continuously and automatically to any point in the Rydberg excitonic series.  If required, the output light could also be frequency doubled using a temperature stabilized periodically poled lithium niobate crystal (ppLN; Covesion MSHG1120-1.0-20) to address the excitons in one-photon excitation.

After the sample, the detection ensemble (blue panel in Fig.\,\ref{Fig:Setup Figure}) was designed to provide complementary scales of spectroscopic resolution, as well as temporal and spatial sensitivity.  For coarse spectral measurements, the signal was passed through a monochromator (Horiba iHR550) with a 2400 lines~\si{\per\milli\meter} grating giving a resolution of 250~\si{\micro\electronvolt}.  Finer spectral details were resolved by a solid temperature-tuned Fabry-P\'{e}rot etalon (FPE; Light Machinery OP-7423-1686-1), which had a free spectral range of $248.6 \pm 0.8$~\si{\micro\electronvolt} and a finesse of $44.5 \pm 0.7$.  Light dispersed by the monochromator or filtered by the FPE was detected by either a CMOS camera (Basler AG ace acA2040-120um) or single photon avalanche detector (SPAD; Micro Photon Devices Srl PDM grade D).  When spectral resolution of the emission was not required, the SHG light was separated from other inelastic emission channels using a bandpass filter (Semrock FF01-580/14-25), and detected using the SPAD. A photon counting card (PicoQuant TimeHarp 260), synchronized with the EOM pulse generator, recorded the relative arrival time of signal photons.  Temporal resolution was ultimately limited by the timing jitter of both the pulse generator driving the EOM and the SPAD, which was $< 50$~\si{\pico\second} in both cases.

The setup also included apparatus for high-resolution imaging of the sample surface.  The sample was illuminated in a K\"{o}hler configuration~\cite{Koehler1893} by light from a yellow LED (Luxeon LXZ1-PX01).  The rear aspheric served as a condenser lens in this configuration and the front aspheric served as the imaging objective.  Additional lenses external to the cryostat completed the microscope.  Using the Basler CMOS as a detector we achieved a resolution of approximately 2~\si{\micro\meter} per pixel over a field of view of $200 \times 200$~\si{\micro\meter}$^{2}$.

\section{Results}

\subsection{Time-resolved measurements and the {two-photon} emission spectrum}

\begin{figure}
    \centering
    \includegraphics[width=85mm]{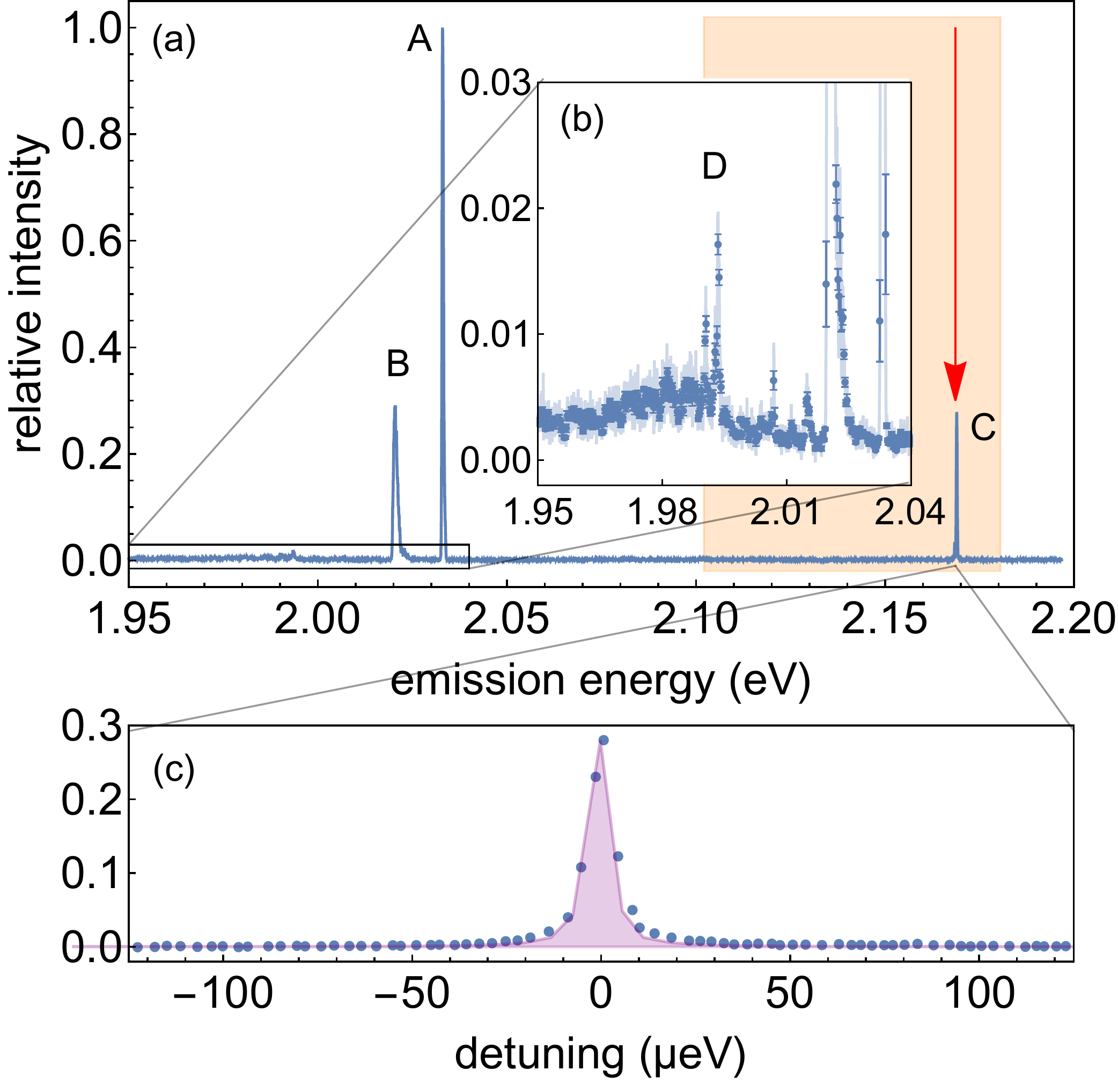}
    \caption{Emission spectrum under TPE resonant with the 5D state.  The excitation pulses were of 1~\si{\nano\second} duration with a peak power of 8.8~\si{\watt} and a repetition rate of 10~\si{\mega\hertz}.  The exposure time was 8 minutes. Three principal features were observed in (a): A: direct quadrupole luminescence from the 1S ortho-exciton; B: $\Gamma^{-}_{3}$ phonon assisted emission from the 1S state and; C: the second harmonic of the laser.  The red arrow indicates the energy of the 5D state.  The orange box represents the 78~\si{\milli\electronvolt} spectral window covered by a bandpass filter used to isolate emission directly from the Rydberg series.  (b) Magnified view of low-intensity light emission around 1.993~\si{\electronvolt}, showing narrow resonances (D) associated with bound exciton states.  (c) Spectrum of the emitted light resolved by the FPE after the bandpass filter (blue data points; error bars smaller than points) compared to that of SHG generated by the ppLN crystal (purple shaded region).}
    \label{Fig:EmissionFigure}
\end{figure}

A typical spectrum of the light emitted by the sample under two-photon excitation is shown in Fig.\,\ref{Fig:EmissionFigure}. Here the excitation laser was two-photon resonant with the 5D exciton level at $2.16879 \pm 0.00002$~\si{\electronvolt}, indicated by the red arrow. There are three dominant features. As in previous studies~\cite{Compaan1972,Petroff1975,Ito1997,Snoke1990,OHara1999,Jang2006,SOLACHECARRANCO2009,Li2013,FRAZER2015,Kitamura2017,Takahata2018}, photoluminescence is observed at the energy of the 1S ortho-exciton recombination line (A) and its $\Gamma^{-}_{3}$ phonon replica (B). These peaks reflect the dominant decay channel of Rydberg excitons, which is phonon-mediated relaxation to the lowest 1S excitonic level, with subsequent radiative decay to the valence band. The strong feature at the energy of the 5D state (C) is the second harmonic of the laser light.  In Fig.~\ref{Fig:EmissionFigure} (c), this light was isolated with the bandpass filter that spanned the entire Rydberg exciton series (orange shaded region in Fig.\,\ref{Fig:EmissionFigure} (a)), before being sent through the FPE. Scanning the etalon revealed a spectral profile that is equivalent to that obtained from laser light frequency doubled in the ppLN crystal, with both measurements limited by the lineshape of the etalon.  The lack of other features in Fig.\,\ref{Fig:EmissionFigure}(c) implies that we do not observe luminescence from the direct radiative decay of  Rydberg states~\cite{Takahata2018,Kitamura2017,Steinhauer2020} within the sensitivity of our instrument.

\begin{figure}
    \centering
    \includegraphics[width=85mm]{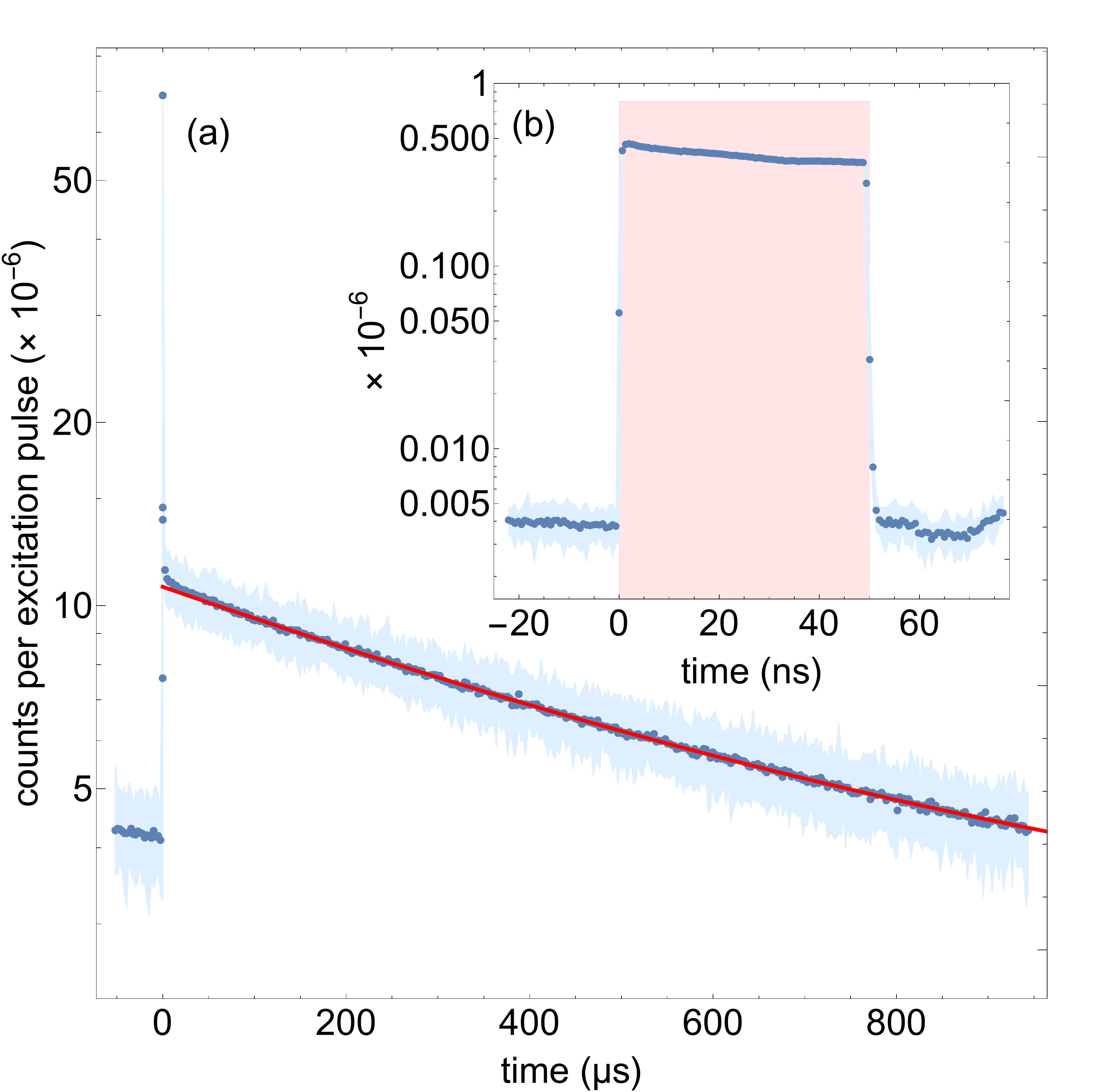}
    \caption{Histograms of photon counts following TPE.  (a) shows the histogram obtained following a 50~\si{\nano\second} excitation pulse resonant with the 9S state at a repetition rate of 1 \si{\kilo\hertz} and peak pulse power of 600~\si{\milli\watt}. The bandpass filter is removed such that all features seen in Fig.\,\ref{Fig:EmissionFigure} contribute. The photon counting card used 51.2~\si{\nano\second} bins, re-binned into 2.5~\si{\micro\second} bins in this figure.  The dark blue data points and the light blue shading represents the mean and range of counts in the new bins respectively.  The solid red line is a fit to the exponential decay with a constant offset described by equation \ref{Eqn:decay}.  The intense spike at $t = 0$ coincides with the laser pulse.  (b) shows a higher resolution histogram obtained with the bandpass filter shown in Fig.\,\ref{Fig:EmissionFigure} in place.  The original bins here were 25~\si{\pico\second} re-binned into 670~\si{\pico\second} bins.  The shaded region indicates the temporal profile of the excitation pulse, which was increased to 1.2~\si{\watt} peak power in order to fill the shorter time bins more quickly. Note that the apparent decay in intensity on the top of this pulse profile is a signature of pile-up~\cite{Davies1970} due to the 77~\si{\nano\second} dead time of the SPAD.}
    \label{Fig:TRHistograms}
\end{figure}

Close examination of the spectrum in Fig.~\ref{Fig:EmissionFigure}(b) also reveals additional sharp peaks (D) at 1.993~\si{\electronvolt}.  To gain further insight, we exploit our pulsed excitation and time-resolved detection system. By using both the EOM and the pulse picker, we probe delays that range over six orders of magnitude from nanoseconds to milliseconds. Figure \ref{Fig:TRHistograms}(a) shows a histogram of the photon counts from all the features in Fig.\,\ref{Fig:EmissionFigure} following a 50~\si{\nano\second} excitation pulse.  Intense emission coinciding with the excitation pulse is apparent in the sharp peak at $t = 0$.  The red fit line traces the steady exponential decay in photon counts, $I(t)$, with the form,

\begin{equation}
    I(t) = I_{0} e^{-t/\tau} + b,
    \label{Eqn:decay}
\end{equation}
where $I_{0}$ is the initial intensity of the emission at time $t=0$ and $b$ is a background count rate.  The lifetime, $\tau$, was measured to be $641 \pm 7$~\si{\micro\second}.  We note that the relative intensities of the features in Fig.\,\ref{Fig:EmissionFigure} (a) are strongly spatially dependent.  A sample location was therefore chosen where the long decay feature in Fig.\,\ref{Fig:TRHistograms} (a) was prominent enough to measure accurately.  We note that this measured lifetime is a remarkably long. For comparison the longest lifetime reported for the lowest-lying 1S para-exciton state is 13~\si{\micro\second}~\cite{Mysyrowicz1979}.  Inserting the bandpass filter suppressed this long-lived component, as shown in  Fig.\,\ref{Fig:TRHistograms}, where the temporal profile of the remaining light traces that of the laser excitation pulse as expected for SHG. We therefore tentatively attribute this lifetime to the narrow resonances observed in Fig.~\ref{Fig:EmissionFigure}(b), since components A and B, which are also blocked by the filter, are associated with the 1S ortho-exciton and are short-lived. Previous detailed studies of the PL spectrum in this energy region revealed the existence of narrow resonances associated with bound excitons~\cite{Jang2006}, that are distinct from the phonon replicas associated with free ortho- and para-excitons~\cite{Takahata2018}. Since these resonances were present in naturally occurring samples, but absent in synthetic material, it was concluded~\cite{Jang2006} that they are associated with unknown metallic impurities. The free 1S para-exciton is predicted to have a radiative lifetime of 7~ms in pure samples~\cite{OHara1999}; our lifetime measurements would therefore be consistent with exciton binding resulting in an enhanced radiative decay rate, as discussed in~\cite{Jang2006}. Lastly we note that the broadband emission underneath the bound exciton peaks observed by~\cite{Jang2006} also appears to be present in Fig.~\ref{Fig:EmissionFigure}(b). 

The results presented in the following sections were obtained with the bandpass filter in place to isolate only the SHG signal; the PL from the 1S exciton and its phonon replicas as well as the long-lived states were filtered out.

\subsection{High-resolution SHG spectroscopy}

\begin{figure*}
    \centering
    \includegraphics[width=170mm]{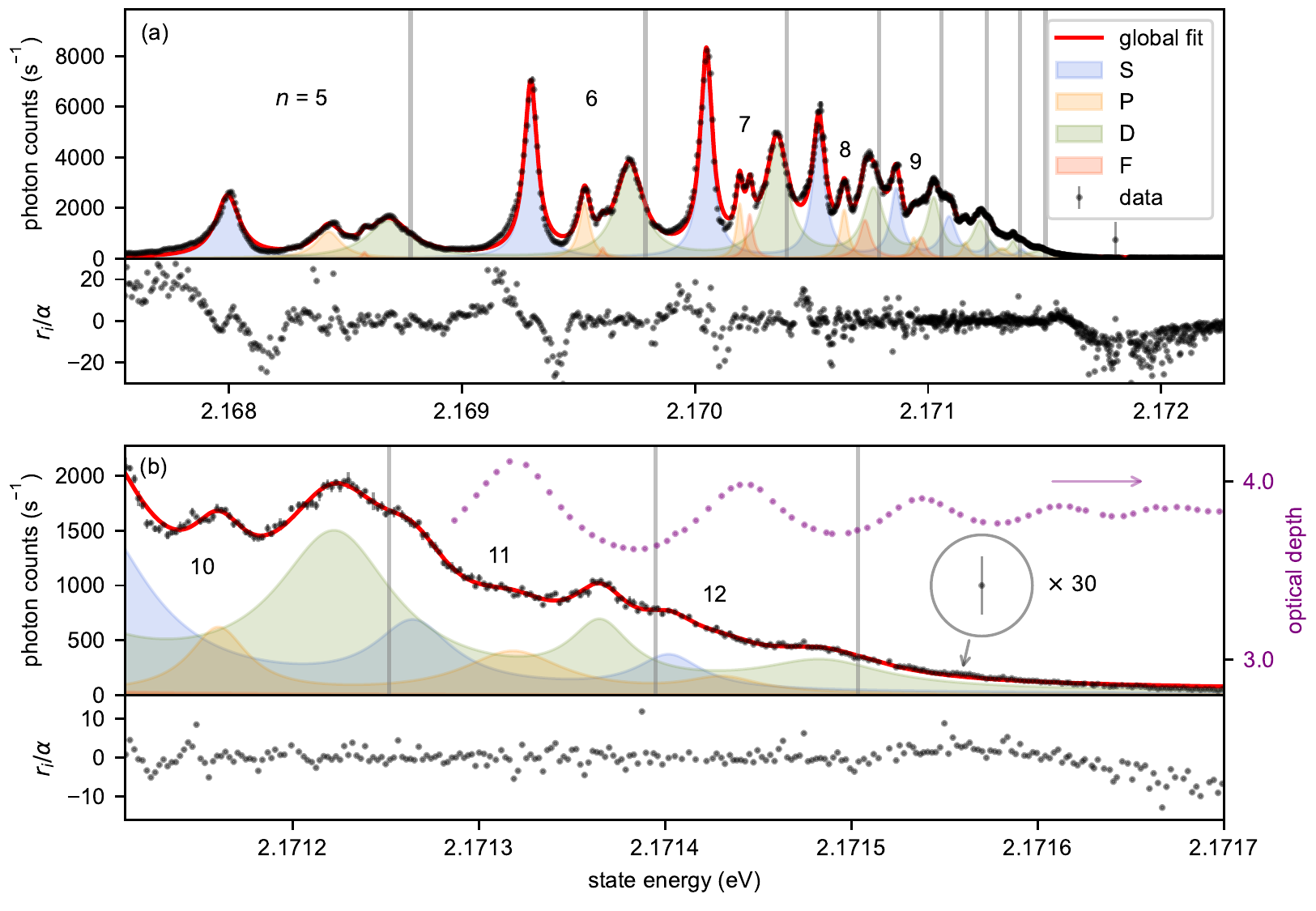}
    \caption{(a) A representative spectrum of the intensity of SHG emitted by cuprite under TPE for $n = 5$ up to the band edge.  (b) shows a zoomed view of the high energy region of (a).  The black data points and their error bars are, respectively, the mean value and standard error of a bin containing 4 measurements at each spectral point.  The error bars are mostly smaller than the data points so a typical data point in (b) is shown with its error bar expanded by a factor of 30.  The red line is a global fit composed of the incoherent sum of Lorentzian lineshapes.  The shaded peaks show the contribution from each angular momentum series.  The lower panels of (a) and (b) show the residuals of the fit normalized to the associated errors.  (b) also shows high $n$ peaks of the P series measured using one-photon absorption in the same sample.  The peaks are annotated with their principal quantum numbers and the vertical gray lines indicate the corresponding point of zero quantum defect.  The excitation consisted of 50~\si{\nano\second} pulses of 160~\si{\milli\watt} peak power at a repetition rate of 5~\si{\mega\hertz}.  The laser was tuned stepwise through the region and the emission light collected for 1~\si{\second} four times per step through a bandpass filter of width 78~\si{\milli\electronvolt} centred at 2.140~\si{\electronvolt}.}
    \label{Fig:SpectralFigure}
\end{figure*}

To obtain a high-resolution exciton spectrum the laser energy was scanned under computer control across the region of interest. For these experiments, the pulse duration was set to 50~\si{\nano\second}, and the repetition rate to 5~\si{\mega\hertz}. The step size was varied between 1.2~\si{\micro\electronvolt}, to capture fine detail at high $n$, to 3.6~\si{\micro\electronvolt} at lower $n$.  At each step, the wavelength was measured and the SHG signal was accumulated on the SPAD four times for 1~\si{\second} each time. Fourteen identical scans were recorded sequentially under the same conditions to produce an estimate of the experimental uncertainties.  An example spectrum, acquired in a single mode-hop free scan across the region $n=5 \rightarrow \infty$ spanning 2.37~\si{\milli\electronvolt} (574~\si{\giga\hertz}) is shown in Fig.\,\ref{Fig:SpectralFigure} (a). A complex series of overlapping resonances extending to $n\approx 12$ is observed. It is apparent that odd-parity $P$ states are observed, in addition to the expected S and D resonances. The amplitude of the peaks has an unusual dependence on $n$, firstly increasing to a maximum at $n = 7$, before decreasing abruptly to zero at higher $n$.

To gain further insight, we fit the entire spectrum with a series of resonances labelled by their principal quantum number $n$ and angular momentum $0\leq L \leq 3$ labelled S,P,D,F respectively. As discussed in previous works~\cite{Uihlein1981,Schweiner2017} these angular momentum labels are only approximate; we  discuss the detailed assignment of the lines in term of the irreducible symmetry representations later.

The measured SHG intensity is given by

\begin{equation}
    I_{\mathrm{SHG}} = A I_{\mathrm{IN}}^2 \left| \sum_{n,L} \chi^{(2)}_{n,L}\right|^2.
    \label{Eqn:CoSum}
\end{equation}
Here $A$ is a scaling factor that includes the detection efficiency and sample thickness, $I_{\mathrm{IN}}$ is the intensity of the IR excitation light, and,

\begin{equation}
    \chi^{(2)}_{n,L}  =  \frac{S_{n,L}}{i \mathit{\Gamma}_{n,l} + (E-E_{n,L})},
\end{equation}
is the contribution to the susceptibility associated with each resonance. The quantity $S_{n,L}$ is a complex amplitude that contains the matrix elements for excitation and emission. 

Fitting the spectrum with equation~\ref{Eqn:CoSum} is an example of an inverse problem; since the intensity is a scalar there is not enough information to unambiguously determine the real and imaginary parts of the amplitudes $S_{n,L}$ without additional information~\cite{Sehmi2017}. Indeed it has been shown~\cite{Busson2009} that for a spectrum containing $N$ resonances, there are $2^N$ sets of parameters that produce identical spectra. The impact of this effect on the parameters (positions, widths) extracted from the  fit is considered in detail in Appendix~\ref{App:Fitting}. To avoid this ambiguity, we use a sum of symmetric Lorentzian profiles

\begin{equation}
    I_{\mathrm{SHG}}= A I_\mathrm{\mathrm{IN}}^2 \sum_{n,L} \frac{\bar{S}_{n,L}^{2}}{\mathit{\Gamma}_{n,L}^{2} + (E - E_{n,L})^{2}}.
    \label{Eqn:IncoSum}
\end{equation}

We note that this incoherent sum neglects the cross terms in equation~\ref{Eqn:CoSum}, and that the resulting amplitudes $\bar{S}_{n,L}$ are real.

As shown in Fig.\,\ref{Fig:SpectralFigure}, the fit based on this Lorentzian lineshape is in very good agreement with the measured data across the full energy range. The experimental uncertainties were weighted by the proximity of the data to the band edge, to reduce the impact of the deviations at low $n$ that dominate the residuals. These deviations are due to a small asymmetry of the S peaks with $n<8$, which is not captured by the model. This asymmetry may be due to residual effects of the interference terms present in equation~\ref{Eqn:CoSum} but neglected in equation~\ref{Eqn:IncoSum}. From this fit, robust values were extracted for the parameters $\bar{S}_{n,L}$, $\mathit{\Gamma}_{n,L}$ and $E_{n,L}$.  Uncertainties on the parameters were obtained from the standard error of the distribution of values produced by fitting each of the 14 spectral scans individually. 

\begin{figure}
    \centering
    \includegraphics[width=85mm]{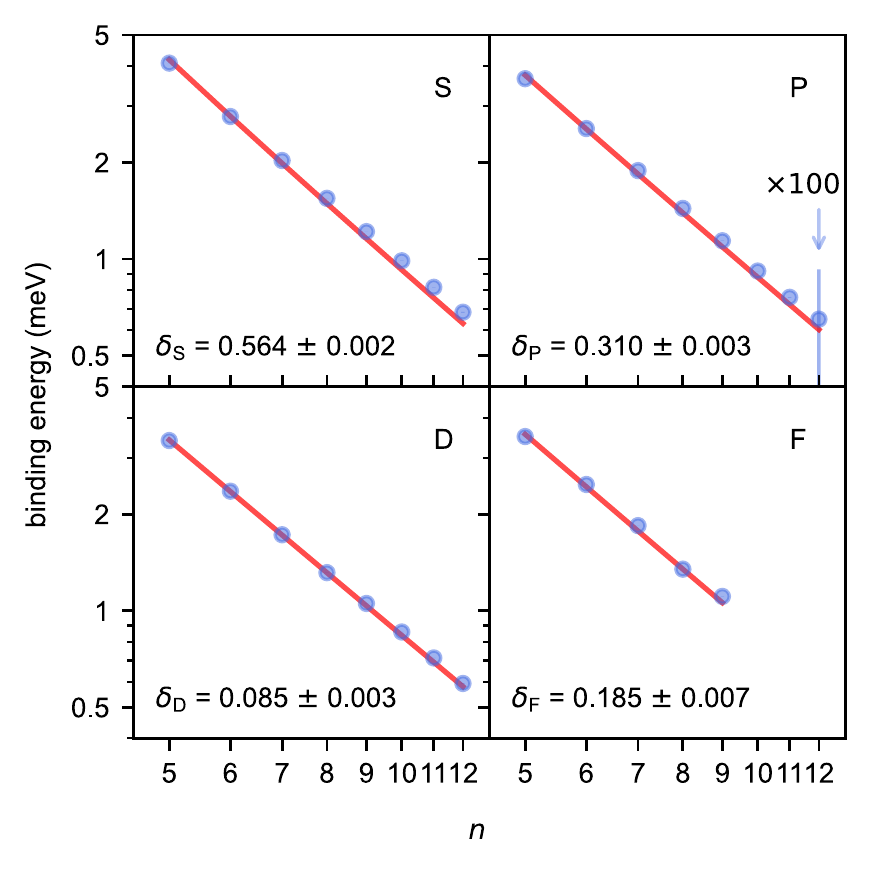}
    \caption{Analysis of the binding energies of the exciton peaks fitted in Figure~\ref{Fig:SpectralFigure}.  One representative error bar has been expanded by a factor of 100 for ease of viewing.  Global values for the bandgap (2.1720780~\si{\electronvolt} $\pm$ 0.4~\si{\micro\electronvolt}) and Rydberg energy (82.40 $\pm$ 0.05~\si{\milli\electronvolt}) were extracted from the fit of all the angular momenta series simultaneously.  These parameters then give rise to the quantum defect for each angular momentum series, $\delta_{L}$.}
    \label{Fig:PeakLocationAnalysis}
\end{figure}

\begin{figure}
    \centering
    \includegraphics[width=\linewidth]{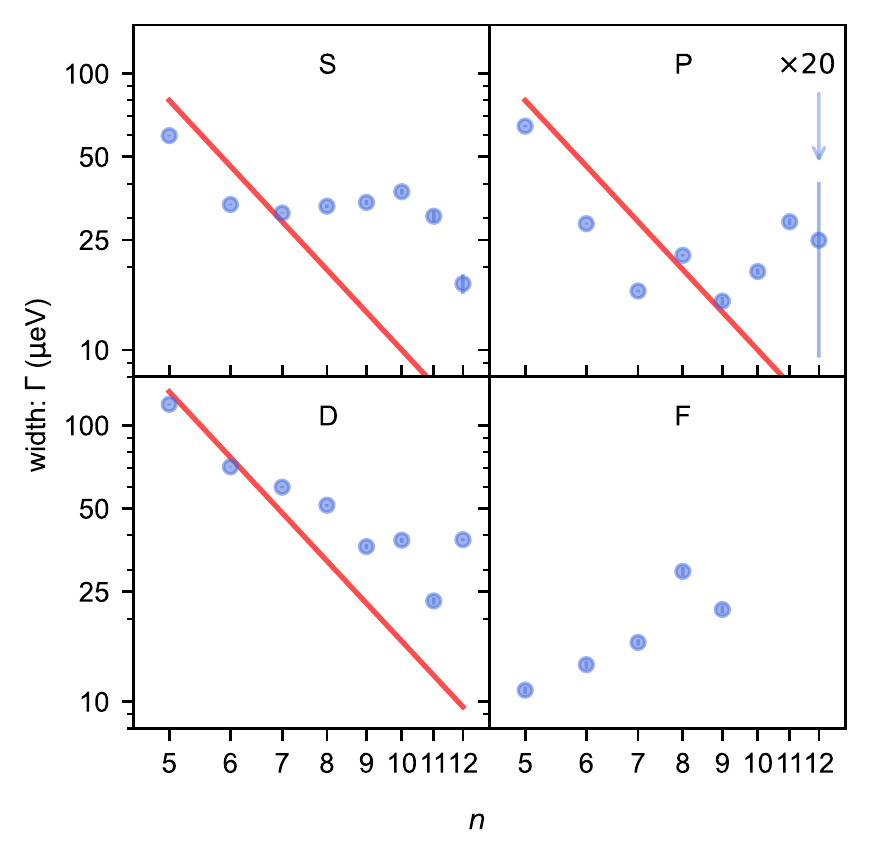}
    \caption{The fitted widths of the excitonic peaks in Figure~\ref{Fig:SpectralFigure} as a function of $n$.  One representative error bar has been expanded by a factor of 20 for ease of viewing.  The red lines represent a $n^{-3}$ trend.}
    \label{Fig:PeakWidthAnalysis}
\end{figure}

\begin{figure}
    \centering
    \includegraphics[width=\linewidth]{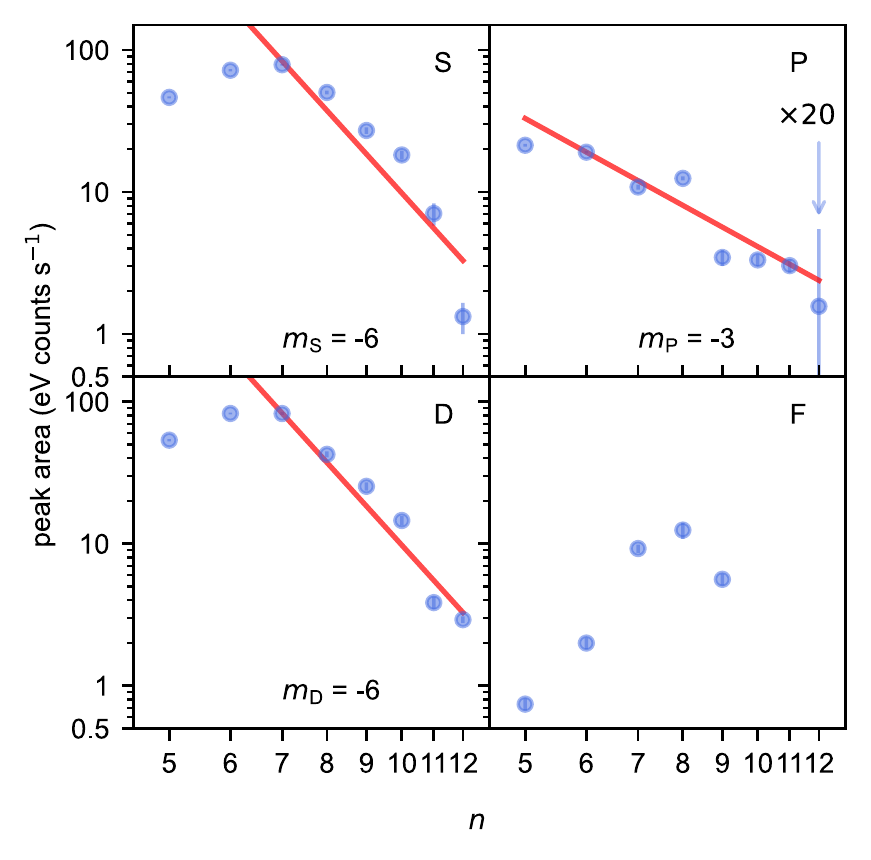}
    \caption{The fitted areas of the excitonic peaks in Figure~\ref{Fig:SpectralFigure} as a function of $n$.  One representative error bar has been expanded by a factor of 20 for ease of viewing.  The red lines are trends fitted to the data ($n > 6$ for S and D) weighted by the corresponding error bars.  The power index $m_{L}$ is indicated for each series.}
    \label{Fig:PeakAreaAnalysis}
\end{figure}

The results for each value of $n$ and $L$ are shown in Figs.\,\ref{Fig:PeakLocationAnalysis} - \ref{Fig:PeakAreaAnalysis}. The resonance energies $E_{n,L}$ (Fig.\,\ref{Fig:PeakLocationAnalysis}) can be described by the Rydberg formula

\begin{equation}
    E_{n,L} = E_\mathrm{g}-\frac{R_{y}}{(n-\mathit{\delta}_{L})^{2}}.
    \label{Eqn:energies}
\end{equation}

where $E_\mathrm{g}$ is the bandgap, $R_{y}$ is the excitonic Rydberg constant and $\mathit{\delta}_{L}$ is the quantum defect~\cite{Kazimierczuk2014,Krueger2020a,Schoene2016}. With the bandgap energy and Rydberg constant constrained to be the same for all four series, fits using equation~\ref{Eqn:energies} are in very good agreement with the resonance positions. These fits yielded  $E_\mathrm{g} = 2.1720780$~\si{\electronvolt} $\pm~0.4$~\si{\micro\electronvolt} and $R_{y} = 82.40 \pm 0.05$~\si{\milli\electronvolt}, and the corresponding $\delta_{L}$ are listed in Table~\ref{Tab:QDefects}.  The values for the bandgap energy and Rydberg constant agree well with those obtained from recent studies of the P odd-parity excitons \cite{Kazimierczuk2014,Versteegh2021}.

Considering the quantum defects, in previous studies, Kr\"{u}ger and Sch\"{o}ne~\cite{Krueger2020a,Schoene2016} measured $\delta_{\mathrm{S}}~\approx~0.5, \delta_{\mathrm{P}}~\approx~0.3, \delta_{\mathrm{D}}~\approx~0.2$, and $\delta_{\mathrm{F}}~\approx~0.1$.  Our values are in reasonable agreement with these previous measurements, except for $\delta_{\mathrm{D}}$, for which we obtain a smaller value.

\begin{table}
    \centering
    \begin{tabular}{c c c c c}
        \hline \hline
        \vspace{-9pt} \\
        $L$ & \hspace{10pt} & Symmetry & \hspace{10pt} & $\delta_{L}$ \\
        \hline
        \vspace{-9pt} \\
        S & & $\mathrm{\Gamma}_{5}^{+}$ & & 0.564 $\pm$ 0.002\\
        P & & $\mathrm{\Gamma}_{8}^{-}$ & & 0.310 $\pm$ 0.003\\
        D & & $\mathrm{\Gamma}_{5}^{+}$ & & 0.085 $\pm$ 0.003\\
        F & & $\mathrm{\Gamma}_{8}^{-}$ & & 0.185 $\pm$ 0.007\\
        \hline
        \hline
    \end{tabular}
    \caption{Quantum defects for each angular momentum series extracted from the global fit in Figure~\ref{Fig:SpectralFigure} and plotted in Figure~\ref{Fig:PeakLocationAnalysis}.  The symmetry representations were obtained from Schweiner~\textit{et al.}~\cite{Schweiner2017}.}
    \label{Tab:QDefects}
\end{table}

At this point it is useful to consider the line assignments in Table~\ref{Tab:QDefects} in more detail. Details of the possible irreducible symmetry representations that may be associated with each angular momentum label $L$ are provided by Schweiner~\textit{et al.}~\cite{Schweiner2017}. 

The $n$D states are shown to consist of a multiplet of states. The $\mathrm{\Gamma}_{3}^{+}, \mathrm{\Gamma}_{4}^{+}$ and the lower energy $\mathrm{\Gamma}_{5}^{+}$ state are close together, split by $< 50$~\si{\micro\electronvolt} at $n = 5$ while the remaining $\mathrm{\Gamma}_{5}^{+}$ state lies 190~\si{\micro\electronvolt} above its low energy counterpart at $n = 5$. This places it above the F multiplet in energy and makes it the highest energy state at a given $n$~\cite{Heckoetter2021}. The value of $\delta_{\mathrm{D}} \approx 0.2$ obtained by Kr\"{u}ger and Sch\"{o}ne~\cite{Krueger2020a,Schoene2016} was obtained from fitting the centre of mass of the multiplet of lines observed in absorption spectroscopy in an applied electric field. In contrast, in our two-photon experiments we are only sensitive to the D states with $\mathrm{\Gamma}_{5}^{+}$ symmetry and in particular, the higher energy state, which is predicted to have a two-photon oscillator strength $\sim 100 \times$ greater than that of its lower energy counterpart~\cite{Schweiner2017}. This explains why we measure $\delta_{\mathrm{D}} < \delta_{\mathrm{F}}$ even though  our values for $\delta_{\mathrm{S}}$, $\delta_{\mathrm{P}}$ and $\delta_{\mathrm{F}}$ are in reasonable agreement with~\cite{Krueger2020a,Schoene2016}.

For the odd-parity states, we note that the $n$P excitons were observed under two-photon excitation in previous studies~\cite{Mund2018}, and that the process has been shown to be weakly allowed~\cite{Mund2019}. Since the location of the P states is well-known from absorption studies, their assignment is straightforward.  The assignment of the weaker F series is more problematic. We cannot rule out that what we have assigned as the F series is in fact the lower energy $\mathrm{\Gamma}_{5}^{+}$ D excitons discussed above.  The two states are very close in energy~\cite{Thewes2015,Schweiner2017,Heckoetter2021} and their oscillator strengths under TPE are both expected to be very small compared to the S and D states, which is borne out in the spectrum.  This assignment remains uncertain without performing experiments under external electric fields.

Next we consider the measured resonance widths, shown in Fig.\,\ref{Fig:PeakWidthAnalysis}. For comparison, we show the predicted $n^{-3}$ trend for non-radiative decay~\cite{Kazimierczuk2014}. For $n>8$ , all the series show significant extra broadening, with the width almost saturating rather than decreasing with $n$. Similar excess broadening for $n>8$ was also observed in one-photon absorption spectroscopy of the $n$P states, even in high quality samples at $< 1$~\si{\kelvin} temperature~\cite{Heckoetter2020}. However, the effect appears much more severe in our spectra, particularly for the S states. The consequence of this saturation is that the peaks begin to merge as they approach each other at high $n$, as is clearly visible in the spectrum.

Lastly we consider the peak areas, which are a measure of the strength of the SHG signal associated with each resonance. The area under each peak shown in Fig.~\ref{Fig:SpectralFigure} was obtained by integration of the fit function. The results are shown in Fig.~\ref{Fig:PeakAreaAnalysis}, where the results are compared to a $n^{-3}$ scaling.

For SHG resonant with the even-parity states, the amplitude $S_{n,L} \propto Q^{nl\rightarrow\mathrm{VB}}M^{\mathrm{VB}\rightarrow nl}$~\cite{Gallagher2021}. Here $Q^{nl\rightarrow\mathrm{VB}}$ is the quadrupole matrix element between the Rydberg state and the valence band, which is predicted to scale as $|Q^{nl\rightarrow\mathrm{VB}}|^2 \propto n^{-3}$~\cite{Schweiner2017,Rommel2021}. The quantity $M^{\mathrm{VB}\rightarrow nl}$ is an effective matrix element for the two photon excitation. It contains products of dipole matrix elements  $D^{\mathrm{VB}\rightarrow\alpha}D^{\alpha\rightarrow nl}$, where $D^{\mathrm{VB}\rightarrow\alpha}$ connects the valence band to the intermediate state $\alpha$, and $D^{\alpha\rightarrow nl}$ connects $\alpha$ to the even-parity excitonic Rydberg state $nl$. Since the dominant intermediate states $\alpha$ are far-off resonant compact states belonging to the blue and violet series, we predict that $D^{\mathrm{VB}\rightarrow\alpha}$ is independent of $n$, while for the second term $|D^{\alpha\rightarrow nl}|^2 \propto n^{-3}$. Taking $\mathit{\Gamma}_{n,L} \propto n^{-3}$ we obtain an overall $n^{-3}$ scaling for the peak area, while for constant width the predicted scaling is $n^{-6}$.

The trends visible in Fig.\,\ref{Fig:PeakWidthAnalysis} are surprising. The data for the {\it even} parity states are in strong disagreement with the predicted scaling, with the area increasing with $n$ to begin with, before dropping much faster than predicted at high $n$. Above $n>12$ no states can be resolved, and the SHG signal goes to zero between 2.1715 -- 2.1717~\si{\electronvolt}. There is a corresponding lump in the residuals in Fig.~\ref{Fig:SpectralFigure} in this region, but no discernible lines. 

At low $n$, the behaviour may be qualitatively understood as a suppression of the amplitude of the lowest $n$ peaks, which is predicted to occur due to state mixing with the the lowest member (1S) of the ``green'' exciton series~\cite{Schweiner2017}. A faster than expected reduction of the oscillator strength for $n>17$ was previously observed in one-photon absorption experiments on the $n$P series in high-quality samples. We compare the behaviour of the even-parity states to that of the $n$P states, observed in both SHG (Fig.\,\ref{Fig:PeakWidthAnalysis}) and using one-photon absorption in the same sample (using the frequency doubled  IR laser, see Fig.\,\ref{Fig:Setup Figure}). In both cases the scaling with $n$ appears more gradual than for the even-parity states, particularly in the one-photon absorption measurements where states up to $n=15$ are visible. 

The apparently abrupt cut-off of the SHG spectrum has important implications for the utility of SHG and the even-parity states in quantum optics applications. The factors governing the visibility of high-$n$ Rydberg excitons have been extensively studied in absorption experiments on the $n$P series~\cite{Lynch2021,Heckoetter2018,Heckoetter2018b,Versteegh2021}, and include Rydberg blockade~\cite{Kazimierczuk2014}, thermally or optically excited free carriers~\cite{Heckoetter2018,Heckoetter2018,Semkat2019}, charged defects~\cite{Lynch2021,Krueger2020,Heckotter2020}, and local strain and disorder~\cite{Heckoetter2018,Versteegh2021}. Therefore to gain further insight we carried out further experiments on the power dependence and spatial variation of the SHG spectrum.

\subsection{Power dependence}

\begin{figure}
    \centering
    \includegraphics[width=85mm]{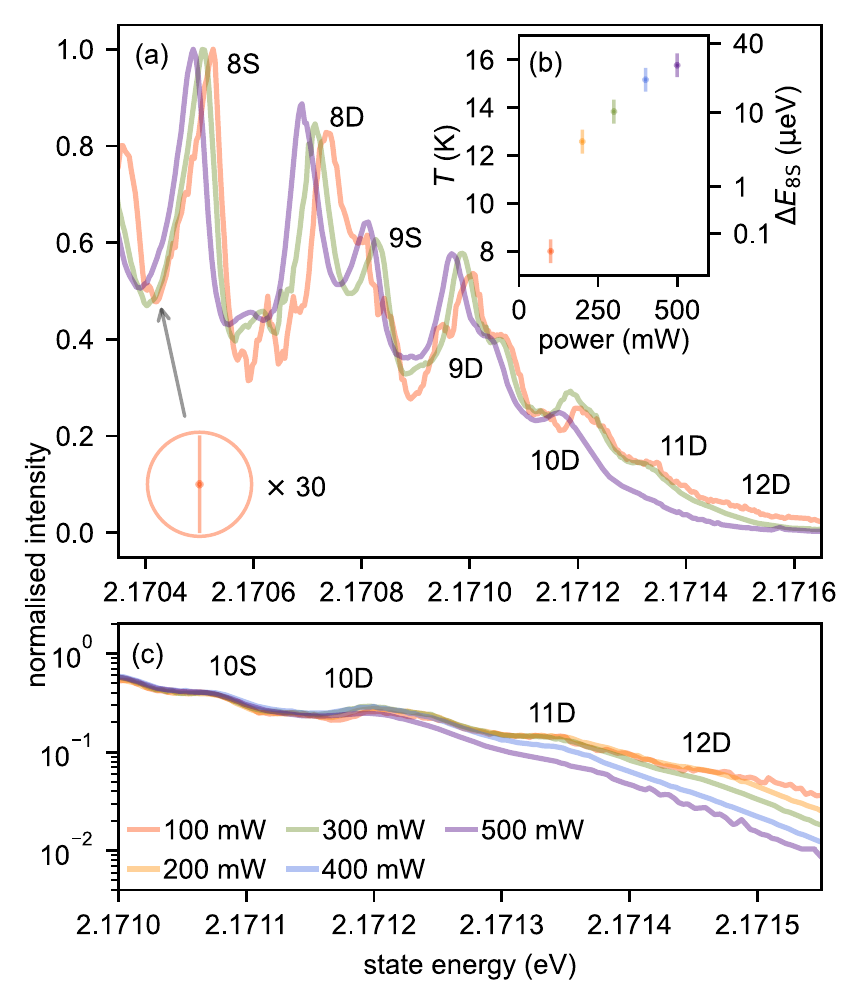}
    \caption{Power dependent excitation spectra. (a) Excitation spectra taken at different excitation powers. 3 of the 5 spectra acquired are shown.  Each has been normalized to the height of the 8S resonance. Spectra show a red shift with increasing power. (b) extracted sample temperature assuming the shift in the 8S resonance is due to the bandgap shift. (c) Zoom of the high $n$ region with the spectral shift applied such that peaks are at the same energy to aid comparison. The logarithmic intensity scale reveals how as excitation power increases, the highest $n$ resonances are no longer resolvable and the SHG intensity vanishes more rapidly.}
    \label{Fig:PowerDependence}
\end{figure}

Power-dependent measurements of the high-$n$ region of the SHG spectrum are shown in Fig.~\ref{Fig:PowerDependence}. By normalizing the intensity of the 8S peak, we remove the quadratic dependence of the SHG intensity on pump power, and reveal the power dependent changes in the shape of the spectrum. We observe a clear shift of the exciton resonances to lower energy with increasing power, as well as an increasing suppression of the high-$n$ peaks.  As shown in~\cite{Heckoetter2018b}, this combination of changes is consistent with increasing temperature, which leads to both a shift of the bandgap and an increased density of thermally excited free carriers. In contrast, non-thermal production of free carriers (by optical excitation or exciton-exciton interactions)~\cite{Heckoetter2018, Heckoetter2018b}, has been shown to lead to a suppression, but not a shift~\cite{Semkat2019}. Using the temperature dependence of the bandgap~\cite{Ito1997}, we can convert the measured energy shifts into a local change in temperature. The result is plotted in Fig.~\ref{Fig:PowerDependence}, where we have constrained the lowest temperature to that obtained from fitting the 1S phonon replica in Fig.~\ref{Fig:EmissionFigure}. These data suggest that the IR light can lead to significant local heating of the sample, which affects the visibility of the highest $n$ excitons. However, while it is clear that there is a significant effect at the highest powers, it is not clear from the data that this effect fully accounts for the observed cut-off of the even-parity spectrum at low power.

\subsection{Spatially resolved two-photon spectroscopy}

Previous experiments have also shown that the visibility of high-$n$ excitons varies locally even in the highest quality samples, with some regions of the sample exhibiting both a stronger lines and a higher maximum value of $n$ \cite{Heckoetter2020}. We have previously shown~\cite{Lynch2021} that our natural samples exhibit variation in the exciton spectrum on a length scale of tens of \si{\micro\meter}. Therefore, we exploited the spatial resolution of our SHG spectroscopy setup to measure the SHG spectrum and the SHG intensity at different locations across the sample.

\begin{figure}
    \centering
    \includegraphics{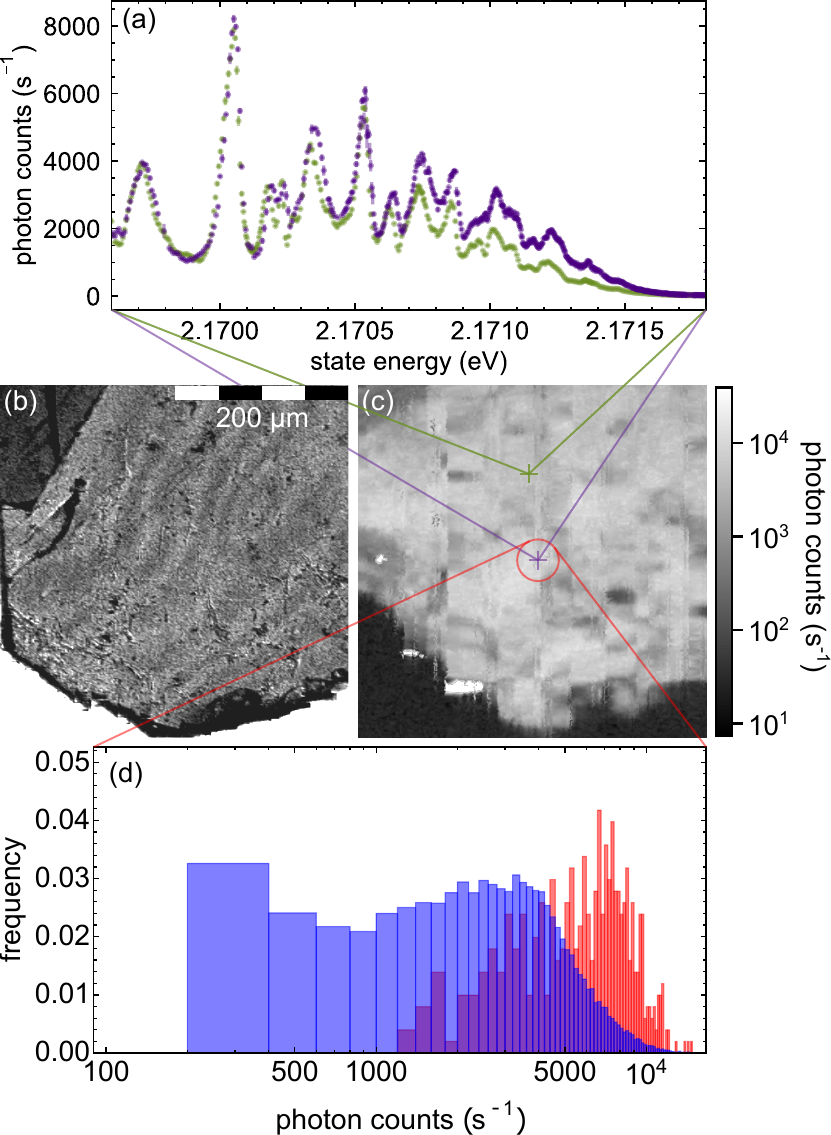}
    \caption{Spatial variation in the sample.  (a) shows two SHG spectra acquired at different locations separated by 100~\si{\micro\meter}.  The sample was excited by 50~\si{\nano\second} excitation pulses of 160~\si{\milli\watt} peak power at a repetition rate of 5~\si{\mega\hertz}.  Images of the sample acquired with the CMOS camera while the sample was under K\"{o}hler illumination from the LED (b) and of SHG emission with the SPAD (c).  The laser was tuned to resonance with the 9D excitonic state and the emission passed through the Rydberg bandpass filter.  (b) and (c) are of the same region of the sample.  (d) shows histograms of photon counts in (c).  The blue data correspond to the whole image and the red data to the small region bounded by the red circle in (c).}
    \label{Fig:SpatialMaps}
\end{figure}

Firstly, we acquired a similar spectrum to that shown in Fig.\,\ref{Fig:SpectralFigure} at a different location in the sample, 100~\si{\micro\meter} away.  This individual spectrum, shown in green in Fig.\,\ref{Fig:SpatialMaps} (a), was very similar to the previous 14 spectral scans at the original location, one of which is shown in purple. The deviations in the location of the resonance positions in the two samples was $< 10$~\si{\micro\electronvolt} for the exciton states with $n > 8$.  Every peak at the second location was slightly red-shifted relative to those at the first location. This confirmed our previous absorption measurements~\cite{Lynch2021} that showed the crystalline environment varying slightly across the sample and giving rise to subtle but measurable changes in peak energies. However, although the amplitudes and widths of the peaks changed dramatically in the spectrum at the new location, the rapid drop to zero of the SHG signal above $n=12$ was still present.

The sample position was then rastered on a 2~\si{\micro\meter} grid by the 3-axis piezo positioning stage to create a spatial map.  A 2D map of the full SHG spectrum across the whole of the sample surface is currently beyond our technical capabilities, due to long-term instabilities in the laser scanning. Therefore to cover more of the sample, we fixed the excitation energy to 2.17102~\si{\electronvolt}, which corresponds to the energy of the 9D state, and measured the SHG intensity as a function of position.  The resulting data are shown in Fig.\,\ref{Fig:SpatialMaps} (c)  along with a visible-light image of the surface of the sample (b). The visible light image reveals the presence of localized voids and inclusions, as well as interference fringes that suggest that the sample is reasonably flat over longer length scales. As shown in Fig.\,\ref{Fig:SpatialMaps} (c) and (d), the SHG intensity varies by over an order of magnitude across the sample. Some of the regions where the SHG is less intense can be correlated with features in (b).  The extremely bright spots at the lower left come from the jagged edge of the crystal and are not included in the histogram. To reduce the effect of these localized imperfections, we zoom in on the 50~\si{\micro\meter} diameter region bounded by the red circle in (c) which appears quite uniform in both Fig.\,\ref{Fig:SpatialMaps} (b) and Fig.\,\ref{Fig:SpatialMaps} (c). Here the histogram was much narrower, although the width of the distribution was much broader than can be accounted for by the measured variation in the 9D line position. In our previous measurements of the spatial variation of the absorption spectrum~\cite{Lynch2021}, the peak energy at $n = 6$ in gemstone material of a similar quality varied by $< 10$~\si{\micro\electronvolt} over a typical 50~\si{\micro\meter} region.  The oscillator strength did not change substantially.  Inspecting Fig.\,\ref{Fig:SpectralFigure}, the corresponding noise in count rate that a peak shift of 10~\si{\micro\electronvolt} would introduce to the SHG signal with the laser tuned near 9D would be on the order of 10\%.  The distribution of counts indicated by the red histogram in Fig.\,\ref{Fig:SpatialMaps} (d) is substantially greater than 10\% from the mode.  Taken together these results suggest that the observed fluctuations in SHG intensity arise primarily from a spatial variation in the SHG susceptibility, rather than merely a shift in the energy of the exciton. Comparable spatial fluctuations were observed previously in SHG arising from the 1S exciton~\cite{Mund2019}, where the authors showed that strain modifies the local SHG selection rules as well as giving rise to an energy shift.  We conclude that SHG spectroscopy is more sensitive to the local sample environment than absorption spectroscopy.  High-resolution imaging is therefore an essential tool to identify regions of good crystal quality as well as sites of potential functional microstructures~\cite{Krueger2018,Konzelmann2019,Steinhauer2020}. 

Concerning the disappearance of high-$n$ states, we note that the same rapid drop-off is observed in both locations probed in Fig.\,\ref{Fig:SpatialMaps} (a), meaning that this behaviour must originate from sample fluctuations on a length scale of $<50$~\si{\micro\meter}. While inhomogeneous broadening due to strain could cause the high $n$ states to be unresolvable, broadening on its own would not cause the observed reduction in oscillator strength. However, if higher $n$ states are more susceptible to strain-induced changes in the selection rules, then local fluctuations in the crystal quality could lead to a more rapid suppression of the high-$n$ peaks in second harmonic generation. 

\section{Discussion}

We have shown that the combination of high-intensity nanosecond pulses and time-resolved single photon detection is a powerful way to study second harmonic generation and related processes in cuprous oxide. In particular, time-resolved measurements revealed the existence of an exceptionally long radiative lifetime which we associate with in-gap resonances at 1.993~\si{\electronvolt}. While bound excitons had been previously observed at this energy~\cite{Jang2006}, this appears to be the first measurement of their radiative lifetime. Future experiments would confirm the assignment of individual resonances to ortho- and para- bound states in~\cite{Jang2006} with spectrally resolved lifetime measurements across the energy range associated with these narrow resonances in Fig.\,\ref{Fig:EmissionFigure}. 

Turning to the main results on second harmonic generation, we note that compared to recent experiments~\cite{Mund2018}, the use of longer bandwidth-limited pulses has enabled direct measurements of the SHG lineshapes without the requirement for a spectrometer. The result is a significant improvement in resolution, and an extension of the observed upper limit for even-parity excitons to $n=12$.  We find that the observed spectrum can be quantitatively fit to a model based on a sum of Lorentzian lineshapes, with possible deviations due to interference terms~\cite{Gruenwald2016} restricted to low-$n$ where individual resonances are well resolved. 

Compared to studies of the odd-parity exciton spectrum using one-photon absorption, we find that the spectrum becomes very crowded at high $n$, with many overlapping resonances. This is partly because there are two strongly allowed even-parity series, in contrast to the odd-parity case where the spectrum is dominated by the P states only. However there is also a strong  contribution from the odd-parity states. As observed in absorption spectroscopy, the resonances broaden faster than the expected $n^{-3}$ trend for $n > 8 $.

A striking feature of the SHG spectrum is the rapid decrease in the intensity of the even-parity peaks for $n>7$, which we do not observe in absorption spectroscopy in the same sample. The observed spatial inhomogeneity of the coupling strength may play a role, as the signal was not collected confocally with the excitation volume and therefore consisted of emission from the full thickness of the sample. However while this effect may wash out the peaks, it doesn't explain why the intensity drops to zero rather than a finite value. Instead our measurements of the power dependence strongly suggest that local heating plays an important role. Empirically we observe that the amount of heating is strongly dependent on the purity of the sample; in samples with a significant concentration of copper vacancies~\cite{Lynch2021}, the heating from below-bandgap absorption of the IR pump light can be sufficient to preclude the observation of SHG.  We must also consider that our back-scattered collection geometry may affect the high-$n$ end of the spectrum.  As the signal light must first pass through the sample before it can be collected, the increasing optical depth near the band edge will suppress that portion of the spectrum to some degree.  A forward collection geometry would be advantageous in reducing the path length of the signal light through the sample.

Regarding the observation of higher principal quantum number, it appears challenging to use SHG to match the $n>25$ observed in absorption spectroscopy of the odd-parity excitons~\cite{Kazimierczuk2014,Versteegh2021}. Nevertheless, our results show there is clear scope for improvement.  Previous studies have suggested that charged impurities set the ultimate limit on the highest observable $n$, as well as contributing a background electric field that causes the anomalous broadening of the resonances~\cite{Heckoetter2020a,Semkat2019,Krueger2020}. Using higher-quality samples with strain-free mounting and careful surface treatment thus appears essential~\cite{Versteegh2021}. Better control of the input polarization would also enable us to fully exploit the selection rules to reduce the intensity of the odd-parity peaks and maximize the SHG efficiency, which in turn would reduce the pump power and the resultant heating.  Two-photon photoluminescence, where the signal is  incoherent emission from the 1S state~\cite{Versteegh2021}, may be more useful than SHG for observing the highest $n$ states as the weak quadrupole emission from the Rydberg state necessary for SHG is eliminated. In future, our experiment can perform simultaneous time-resolved measurements of both PL and SHG with appropriate filters, allowing these techniques to be compared. Lastly we note that the highly coherent, narrowband nature of the SHG produced in our experiment makes it very useful for applications, even within the currently observed range of principal quantum numbers. An example is the recent observation of the coherent modulation of the second harmonic by an applied microwave electric field~\cite{Gallagher2021}.

\section{Conclusion}

We have carried out high-resolution laser spectroscopy of the Rydberg excitonic states of cuprous oxide using second harmonic generation. We have assigned even-parity excitonic states up to  $n=12$ and measured their linewidth for the first time.  We have also described a versatile experimental platform for studying the properties of Rydberg excitons in cuprous oxide, which has sufficient spectral, temporal and spatial resolution to study the SHG process in detail, and to reveal the existence of extremely long-lived bound exciton states.

\section{Acknowledgement}

This work was supported by the Engineering and Physical Sciences Research Council (EPSRC), United Kingdom, through research grants EP/P011470/1 and EP/P012000/1. The authors also acknowledge seedcorn funding from Durham University. LG acknowledges financial support from the UK Defence and Scientific Technology Laboratory via an EPSRC Industrial Case award.  We are grateful to Ian Chaplin and Sophie Edwards (Durham University, Department of Earth Sciences) for the slicing and polishing of the samples used in this work, and to Patric Rommel for discussions of the scaling laws for quadrupole emission. Information on the data underpinning the results presented here, including how to access them, can be found in the Durham Research Online repository at doi.org/10.15128/r1t435gd034.

\bibliography{manuscript}

\section{Appendix}

\begin{table*}
    \centering
    \begin{tabular}{c c c c c}
        \hline \hline
        \vspace{-9pt} \\
        $n$ & S & P & D & F \\
        \hline
        \vspace{-9pt} \\
        5 & 2.1679940(3) & 2.1684260(4) & 2.1686815(4) & 2.1685826(9) \\
        6 & 2.1692950(4) & 2.1695251(2) & 2.1697171(3) & 2.1696006(5) \\
        7 & 2.1700486(2) & 2.1701904(3) & 2.1703514(2) & 2.1702344(4) \\
        8 & 2.1705313(2) & 2.1706394(3) & 2.1707626(9) & 2.1707290(1) \\
        9 & 2.1708600(4) & 2.1709375(4) & 2.1710237(2) & 2.1709689(7) \\
        10 & 2.1710903(5) & 2.1711604(3) & 2.1712190(11) & \\
        11 & 2.1712615(10) & 2.1713181(11) & 2.1713654(6) & \\
        12 & 2.171397(2) & 2.171430(3) & 2.1714848(11) & \\
        \hline
        \hline
    \end{tabular}
    \caption{Energies of the $|n,L\rangle$ states presented in Fig~\ref{Fig:PeakLocationAnalysis}.  These are the average and standard error of the values extracted from fitting the incoherent sum of Lorentzian peaks to the 14 individual spectral scans.}
    \label{Tab:StateEnergies}
\end{table*}

\subsection{Fitting with complex poles} \label{App:Fitting}

In this Appendix we consider the function used to fit the SHG spectrum in Fig.\,\ref{Fig:SpectralFigure} in more detail. Specifically, we analyze the approximation of the expected expression for the SHG intensity, that is the sum of the square of the sum of complex poles given in equation~\ref{Eqn:CoSum}, by the sum of Lorentzians given in equation~\ref{Eqn:IncoSum}.

 \begin{figure*}
    \centering
    \includegraphics[width=170mm]{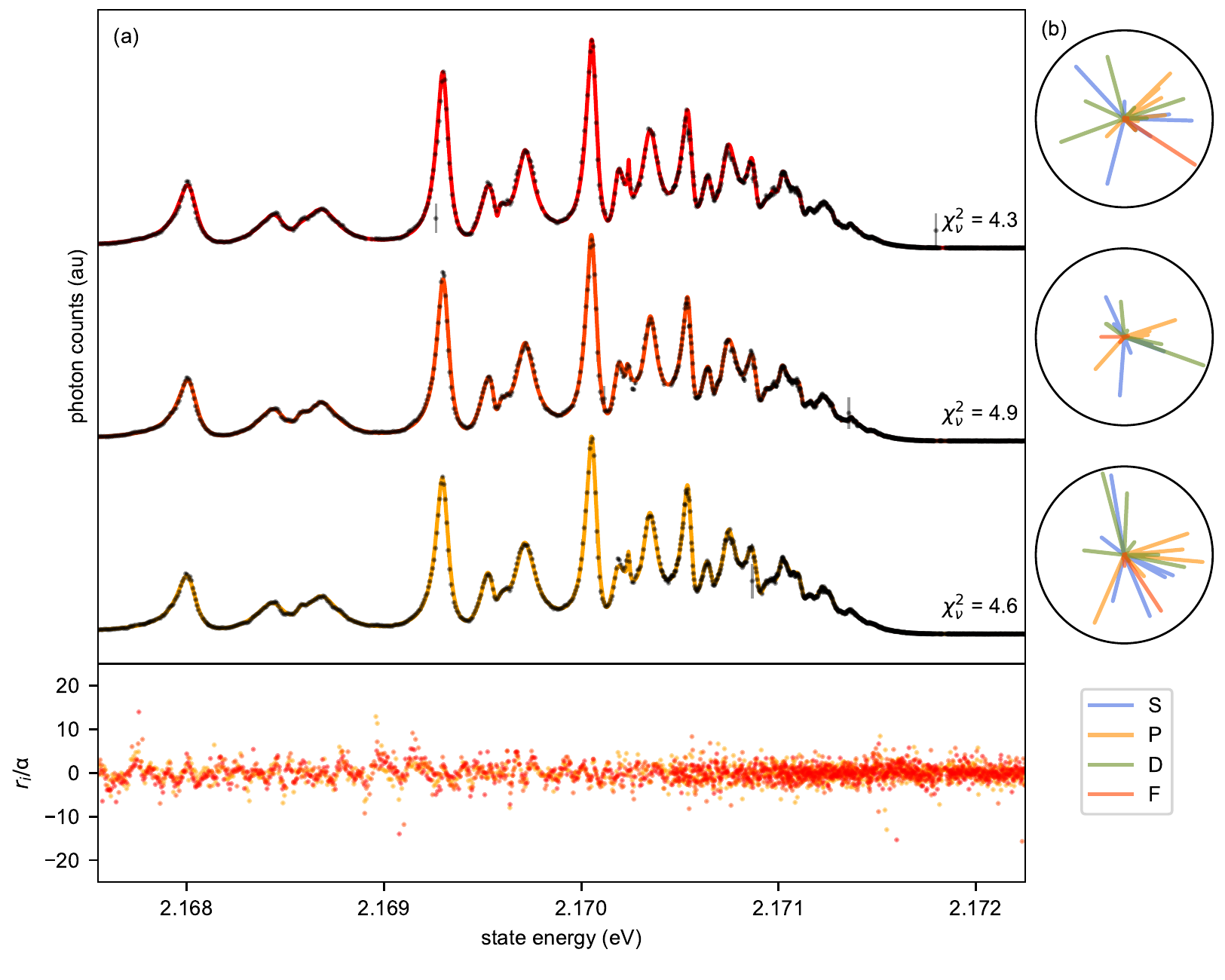}
    \caption{Fits to spectral scans using the coherent sum of complex poles.  (a) Three representative scans measured under identical conditions plotted with their associated global fit functions.  Each fit is annotated with its reduced $\chi^{2}$.  The bottom panel shows the normalized residuals colour coded according to that of the fit line above.  (b) Modulus and argument of the complex amplitude of each pole represented by phasor diagrams.  The associated angular momentum assignment $L$ of each pole is colour coded according to the legend.}
    \label{Fig:PhasorPlot}
\end{figure*}

The central issue is that the intensity alone does not provide sufficient information to unambiguously define the amplitude and relative phase of each pole in equation~\ref{Eqn:CoSum}.  To demonstrate the problem, Fig.\,\ref{Fig:PhasorPlot} provides examples of how  equation~\ref{Eqn:CoSum} may yield fits that are indistinguishable in terms of their residuals and goodness-of-fit, but which provide radically different values for the real and imaginary parts of the amplitudes of each pole. The use of the incoherent sum of Lorentzians (equation~\ref{Eqn:CoSum}) in Fig.\,\ref{Fig:SpectralFigure} of the main text avoids this problem, at the cost of forcing the amplitude to be real. 

It is instructive to compare the Lorentzian fit in Fig.\,\ref{Fig:SpectralFigure}, and the fits using the sum of complex poles in Fig.\,\ref{Fig:PhasorPlot}. Overall the Lorentzian fit is worse, with a reduced $\chi^{2}$ value of $\chi^{2}_{\nu} = 118.2 \pm 1.6$ over the 14 scans, rather than $\chi^{2}_{\nu} = 4.9 \pm 0.2$. Comparing the residuals, it is clear that this difference is due to an asymmetry of the 5S, 6S and 7S lines, that cannot be reproduced by the symmetric Lorentzian lineshape, but which is captured by the sum of complex poles. Over the rest of the spectrum, the quality of the two fits is essentially indistinguishable. The phasor plots in  Fig.\,\ref{Fig:PhasorPlot} provide an explanation for this behaviour. At high $n$ where the states overlap, the essentially random distribution of the phasors obtained from each of the three fits yields a total phase that averages to zero. The cross terms in equation~\ref{Eqn:CoSum} therefore average out. At lower $n$ where the states are more clearly resolved the fit is more sensitive to these relative phases.

\begin{figure}
    \centering
    \includegraphics[width=\linewidth]{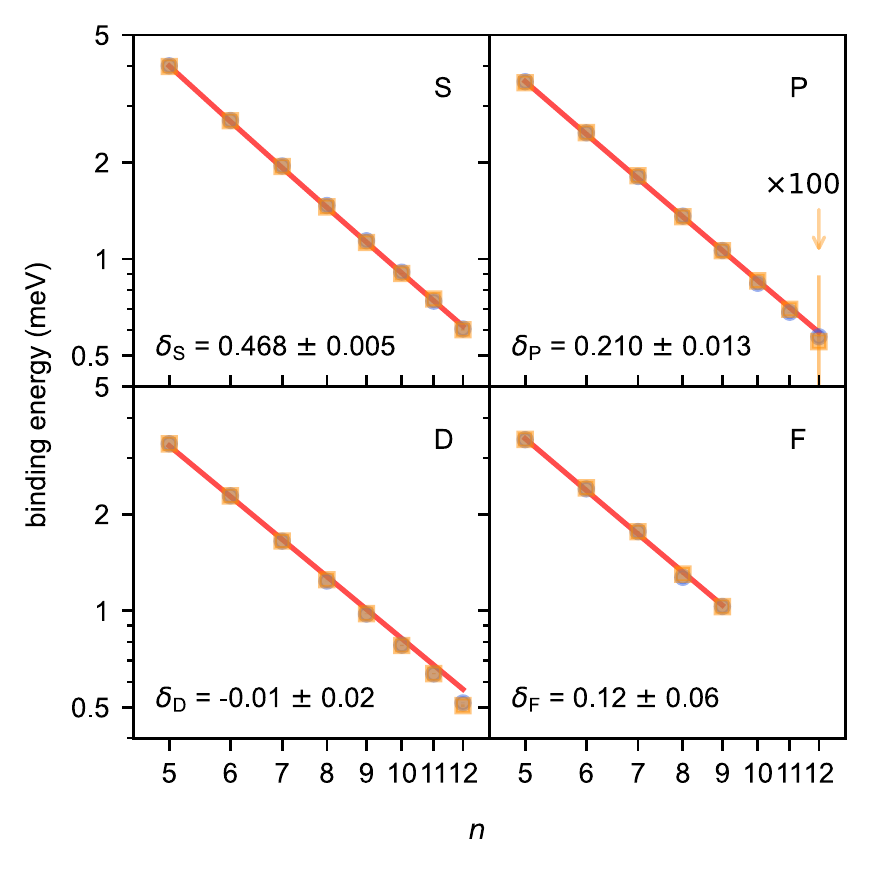}
    \caption{The fitted positions of the excitonic peaks in the 14 spectral scans as a function of $n$ by the two different methods.  The red lines represent the global fit to the Rydberg formula in Eq.~\ref{Eqn:energies}.  Blue circles correspond to the incoherent sum of Lorentzians (also presented in Fig.\,\ref{Fig:PeakLocationAnalysis}) and the orange squares to the coherent sum of complex poles.  The quantum defects annotated are obtained from the coherent model.}
    \label{Fig:ComparingPeakEnergies}
\end{figure}

\begin{figure}
    \centering
    \includegraphics[width=\linewidth]{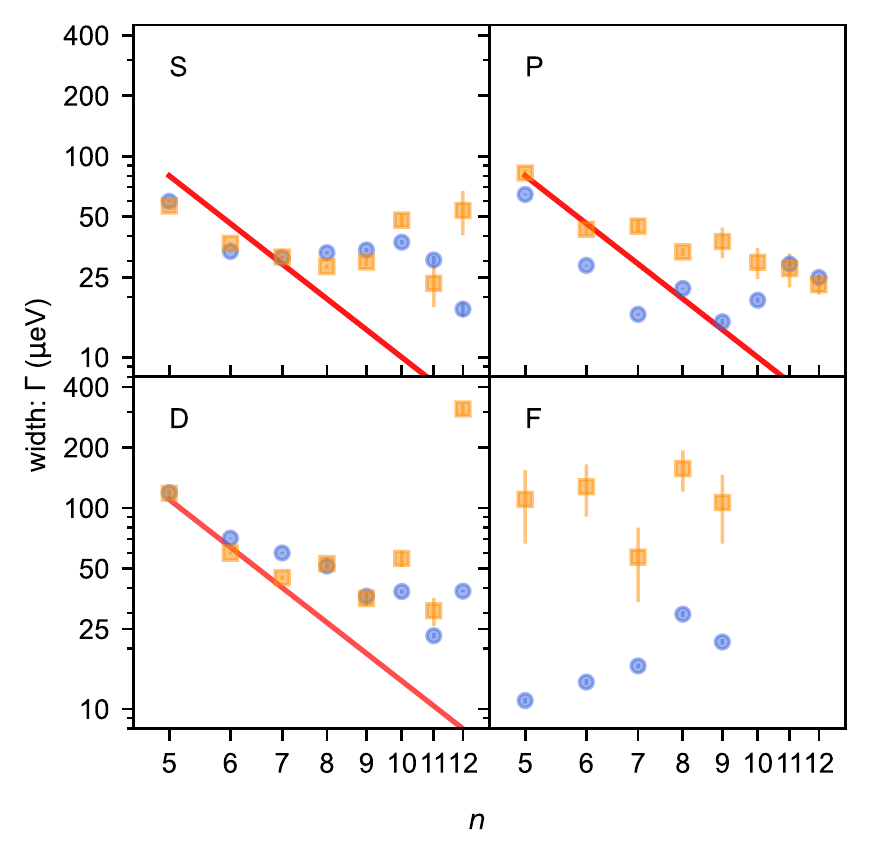}
    \caption{The fitted widths of the excitonic peaks in the 14 spectral scans as a function of $n$ by the two different methods.  The red line indicates the $n^{-3}$ trend measured for the P series by Kazimierczuk \textit{et al.}. Blue circles correspond to the incoherent sum of Lorentzians (also presented in Fig.\,\ref{Fig:PeakWidthAnalysis}) and the orange squares to the coherent sum of complex poles.}
    \label{Fig:ComparingPeakWidths}
\end{figure}

Lastly for completeness we compare the peak locations and widths obtained from the two methods in Figures Figs.~\ref{Fig:ComparingPeakEnergies} and~\ref{Fig:ComparingPeakWidths}.  Overall, the peak locations are in good agreement between the two methods.  The uncertainties on the quantum defects are between 2 and 10$\times$ larger from the coherent model than those from the incoherent model, reflecting the small energy shifts that the random phases impart to the centre of the peaks.  The bandgap and Rydberg energy from this treatment are $E_\mathrm{g} = 2172000.7 \pm 1.4$~\si{\micro\electronvolt} and $R_{y} = 82.2 \pm 0.2$~\si{\milli\electronvolt} respectively.  These values and those of the quantum defects are close to those of the incoherent model but do not agree within the corresponding uncertainties. This discrepancy probably reflects the method used to calculate the uncertainty in the parameters, which considers the statistical distribution of the fit parameters over multiple datasets, but neglects systematic errors in the fit. The widths of the excitonic states extracted by the two models are shown in Fig.\,\ref{Fig:ComparingPeakWidths}.  The red lines show the $n^{-3}$ dependency Kazimierczuk \textit{et al.} measured in the absorption spectrum.  The values are generally comparable between the two models but differ for individual states by up to a factor of 5, particularly for the F states. The uncertainty in the width of each state extracted from the statistical distribution of the fits over the 14 datasets is significantly larger in the case of the complex poles fit, as expected. There is no well-defined relationship with $n$ in either model but both exhibit a general trend to narrower peaks at higher $n$, except for the case of the  F series.

\end{document}